# Ion and water transport in charge-modified graphene nanopores[*]

Yinghua Qiu, Kun Li, Weiyu Chen, Wei Si, Qiyan Tan, and Yunfei Chen[†]

*School of Mechanical Engineering and Jiangsu Key Laboratory for Design and Manufacture of Micro-Nano Biomedical Instruments, Southeast University, Nanjing, 211189, China*

**Abstract：**

Porous graphene has high mechanical strength and atomic layer thickness, which make it a promising material for material separation and biomolecule sensing. Electrostatic interactions between charges in aqueous solution are a kind of strong long-range interaction which may have great influence on the fluid transport through nanopores. Here, molecular dynamics simulations were conducted to investigate ion and water transport through a 1.05-nm-in-diameter monolayer graphene nanopore with its edge charge-modified. From the results, it is found that the nanopores are selective to counterions when they are charged. As the charge amount increases, the total ionic currents show an increase-decrease profile while the co-ion currents monotonously decrease. The co-ions rejection can reach 75% and 90% when the nanopores are negatively and positively charged, respectively. $Cl^-$ ions current increases and reaches a plateau, and $Na^+$ current decreases with the charge amount in the systems they act as counterions. Besides, the charge modification can enhance the water transport through nanopores obviously. This is mainly due to the ion selection of nanopores. Especially, positive charges on the pore edge facilitate the water transport much more than negative charges.

[*]Project supported by the National Basic Research Program of China (Grant Nos. 2011CB707601, 2011CB707605), the National Natural Science Foundation of China (Grant No. 50925519), the Fundamental Research Funds for the Central Universities, Funding of Jiangsu Innovation Program for Graduate Education (Grant No. CXZZ13_0087), and the Scientific Research Foundation of Graduate School of Southeast University (YBJJ 1322).

[†]Corresponding author. E-mail: yunfeichen@seu.edu.cn

**Key words：**

Monolayer porous graphene, ion selection, charge-modified nanopore, water transport, ionic current

**Introduction：**

As a single layer two-dimensional crystal constituting of carbon atoms,[1] graphene has deserved lots of attention in recent years. Pristine graphene has excellent thermal property[2] and remarkable electronic property,[3] which make it widely used in thermal devices,[4] and energy storage such as supercapacitors[5] and Li-ion batteries.[6]

Simultaneously, monolayer graphene is extremely thin, the thickness of which can reach atom scale, but has outstanding mechanical properties.[7] Nanopores can be drilled in graphene membranes by transmission electron microscope (TEM)[8] or other methods.[9, 10] With these pristine nanopores, the reduced porous graphene layer becomes indispensable for material separation, such as gas selection,[9, 11] and biomolecules sensing such as DNA sequencing.[8, 12] These applications mainly take advantage of the size excluding effect of nanopores, that is, small molecules can pass through nanopores while larger ones cannot. Due to its relative inertness and impermeability to all standard gases, graphene provides an excellent material for molecular sieves. For example, with ultraviolet-induced oxidative etching, Koenig *et al.*[9] created sub-1-nm pores in micrometer-sized graphene membranes. Based on the sizes of induced pores, their resulting membranes could be used to select specific gas molecules, such as $H_2$ and $CO_2$. Since its thickness is only one carbon atom diameter, which is comparable to the distance between base pairs,[13] monolayer graphene is a promising candidate for DNA sequencing. Recently, lots of simulations and experiments have been conducted through measuring the ionic current accompanying DNA molecules translocation across the nanopores.[8, 12] However, their results showed ionic current noise levels are several orders of magnitude larger than those in silicon nitride nanopores[12] which made it difficult to discriminate the four kinds of bases.

The applications stated above are based on the pristine porous graphene membranes. In order to expand the function of graphene layers, lots of graphene derivatives are found.[14] Graphene oxide (GO) is a most important one which is easy to fabricate and should be amenable to industrial-scale production. Due to its advantages such as hydrophilicity, GO layer facilitates the material separation, especially about water molecules separation in vapor or liquid phase. Nair *et al.*[15] found that submicrometer-thick membranes made from GO could be completely impermeable to liquids, vapors, and gases, but these membranes allow unimpeded permeation of water. They attributed these seemingly incompatible observations to a low-friction flow of a monolayer of water through two-dimensional capillaries formed by closely spaced graphene sheets. Joshi *et al.*[16] investigated permeation through micrometer-thick laminates prepared by means of vacuum filtration of GO suspensions. The laminates act as molecular sieves, blocking all solutes with hydrated radii larger than 4.5 angstroms. Smaller ions permeated through the membranes at rates thousands of times faster than what is expected for simple diffusion which is attributed to a capillary-like high pressure acting on ions inside graphene capillaries.

Besides the derivatives, graphene can be functionalized with different chemical groups by various means.[17, 18] These functional ligands can realize or facilitate the filter functions of pristine graphene membranes. Many researches on modified porous graphene have been conducted which provide inspirations to the design of nano-devices based on porous graphene, particularly in desalination. With molecular dynamics (MD) simulations, Cohen-Tanugi *et al.*[19] pressed brine through graphene nanopores to measure the desalination performance as a function of pore size, chemical functionalization, and applied pressure. Their results indicated that the membrane's ability to prevent salt passage depended on the kinds of chemical functional group besides pore size and pressure. Hydroxyl groups could roughly double the water flux due to their hydrophilic character. Based on the ion prevention and enhanced water flux, chemical functional groups modified nanopores could be used in sea water desalination with lower energy cost.[20] Sint *et al.*[21] designed graphene nanopores functionalized by negatively charged nitrogen and fluorine as well as positively charged hydrogens, respectively. They found that the functionalized nanopores had selection to the ions with counter-charge from the function groups. Through analyzing the potential of mean force along the direction perpendicular to the pore, Konatham *et al.*[22] detected the ease of ion and water translocation across modified nanopores with different chemical groups in NaCl solution. They found that charge groups could exclude co-ions effectively, but the effect became less pronounced as the pore diameter increased.

These chemical groups tethered on the rim of the nanopore mainly change the charge property of the nanopores except some groups can change the pore hydrophobicity. Because the electrostatic force in solution is a strong long-range interaction,[23] the surface charges may have great influence on the nanopore selection to ions, as well as the water transport, which is particularly important in sea water desalination. With MD simulation, we investigate the ion and water transport in a 1.05 nm nanopore to detect of influence of the charge sign and amount on their translocation across nanopores. The pore size in the simulation is moderate due to the smaller size effect of pore[16, 22] i.e. ions and water molecules can pass through and the charges on the edge have obvious effect on the transport of solution. As an effective tool to discover microscopic characteristics of material, MD simulation provides detailed characteristics of nanofluid which facilitate us understand this process and mechanism of nanopore selection to ions. We think this may provide some useful information to the device design for desalination and other domains involving porous graphene.

**MD simulation detail：**

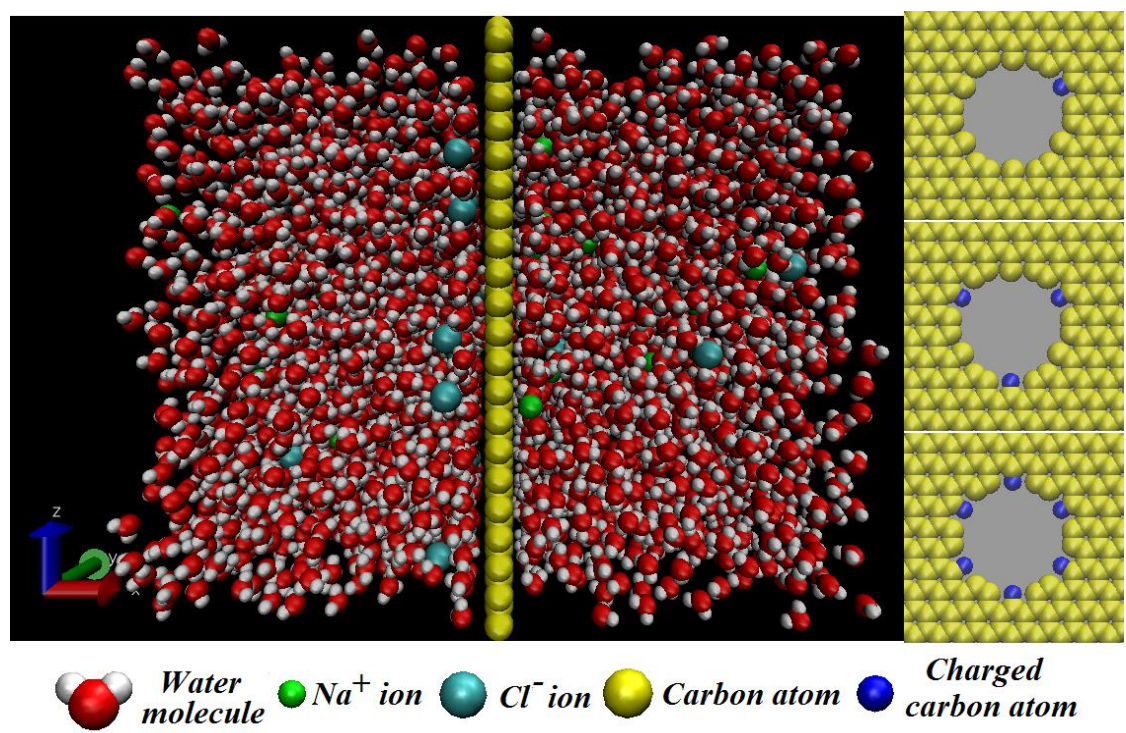

Fig. 1. (color online) Schematic diagram of the MD model.

The schematic diagram of the model used in the simulations is exhibited in Fig. 1,[24] which is assumed to be infinite in all the three directions using periodic boundary conditions. The graphene layer with a nanopore is set in the center of the system. The height and the width of the box are 5.0 nm and 2.84 nm, respectively. On each side of the graphene layer, solution extends 3.2 nm along the axial direction of the nanopore. The edge of the pore is charged and the charge atoms are distributed as shown on the right of Fig. 1. In our simulation, an external electric filed with strength 0.25 V/nm was applied along minus *x*-direction. During the simulation, the carbon atoms of graphene layer were frozen without thermal vibration. In order to realize the neutrality of the systems, extra counterions were added in the systems. The numbers of water molecules used in the simulations are 2880 and the concentration of NaCl is 0.5 M.

In our simulations, the TIP4P model[25] was selected to simulate the water molecules and the SETTLE algorithm[26] was chosen to maintain the water geometry.

The Lennard-Jones (LJ) potential was used to describe the interactions between different atoms, except interaction pairs involving hydrogen atom and carbon-carbon pairs.[19] The electrostatic interactions among ions, water molecules and pore charges were modeled by Ewald summation algorithm.[27] The motion equations were integrated by the leap-frog algorithm with time step of 2.0 fs. The solution system was maintained at 298 K by Berendesen thermostat.[28] The first run lasting 1.5 ns was used to equilibrate the system. Another 2-ns-long run was followed to gather the statistical quantities.

**Results and discussions**

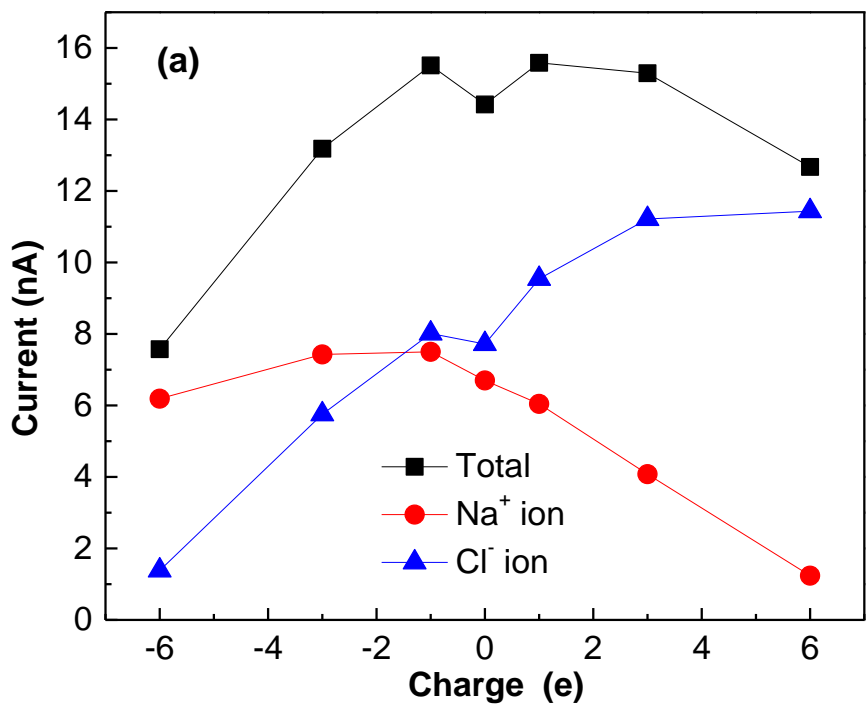

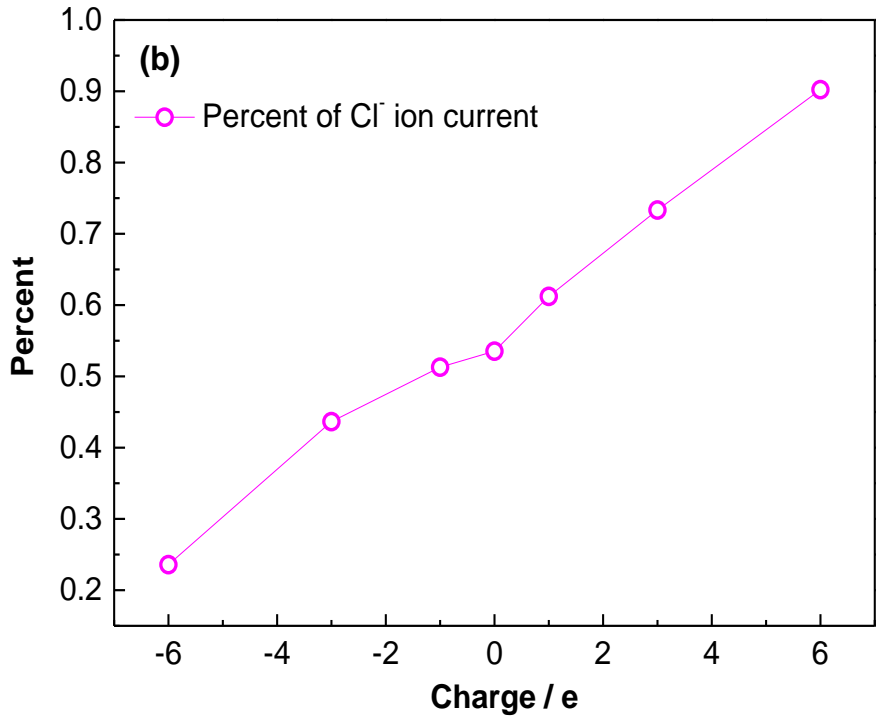

Fig. 2. (color online) (a)The current through the nanopores with different charges. (b) The percent of $Cl^-$ ion current to the total current with pore charges changing. $e$ is elementary charge. The positive and negative numbers on the horizontal axis correspond to positive and negative charges on the pores.

Figure 2(a) shows the ionic current as a function of the sign and amount of charges on the nanopore edge. In the neutral nanopore, $Cl^-$ ion current is a little larger than $Na^+$ ion current due to the higher mobility of $Cl^-$ ions.[29] When the nanopore is charged, the total current shows an increase-decrease profile. When the charges are negative, the modified nanopore provides an attractive force to $Na^+$ ions and repulsive force to $Cl^-$ ions, which results the nanopore selection to $Na^+$ ions. As the charge amount increases, both $Na^+$ ion and $Cl^-$ ion currents decrease. This is attributed to the enhanced prevention of $Cl^-$ ions and attraction to $Na^+$ ion when they have passed through the pore (shown in the following). However, when the nanopore has positive charges, $Cl^-$ ions act as counterions and main current carriers. The nanopore has a specific selection to $Cl^-$ ions. With charge amount increasing, $Cl^-$ ion current increases monotonously and reaches a plateau accompanying with the total current drop which makes the percent of Cl¯ ion current to the total current much larger as shown in Fig.2 (b). As the pore charges change from −6$e$ to 6$e$, the percent of Cl¯ ion current rises monotonously from 23.5% to 90.2%, which means the co-ions rejection by the negatively and positively charged nanopores can reach 76.5% and 90.2%, respectively.

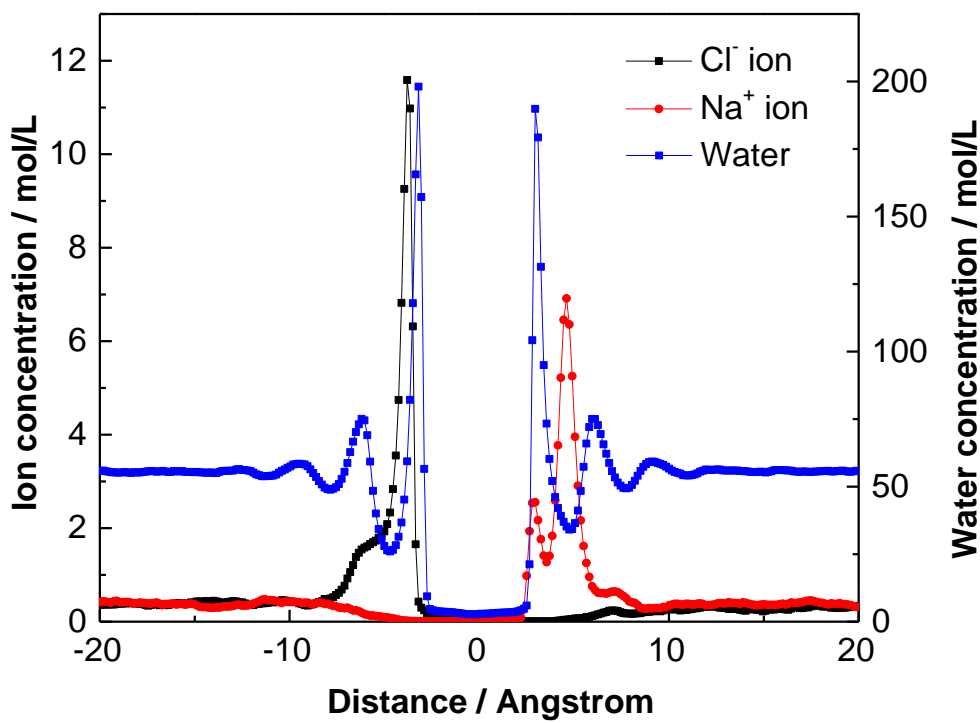

Fig.3. (color online) The ion and water distributions perpendicular to the neutral porous graphene.

When the system reached equilibrium under the electric field, ion and water concentrations perpendicular to the graphene layer were analyzed. The results obtained in neutral case are plotted in Fig. 3. The monolayer graphene nanosheet was set at the zero position. When an external electric filed is applied, Na$^+$ and Cl¯ ions are driven directionally and accumulate on the right and left side of the graphene layer respectively. The accumulating ions show layering effect in the figure. Their main peak locations of Na$^+$ and Cl$^-$ ions are 4.6 and −3.8 Å. The difference is that the smaller peak of Na$^+$ ions profile is nearer to the graphene layer, while that of Cl$^-$ ions is farther away. The distribution shows that a few Na$^+$ ions stay near the graphene which are attracted by the water molecules adsorbed by the membrane. By comparing the peak locations of ions and water, the most ions accumulate in the valley of water concentration which determines the main peak locations of ions. For water density profile, the adsorption layer positions on each side of the graphene are not symmetrical i.e. the left peak is a little farther than the right one. This is mainly caused by the different structures of water molecules in adsorption layers under the electric field.

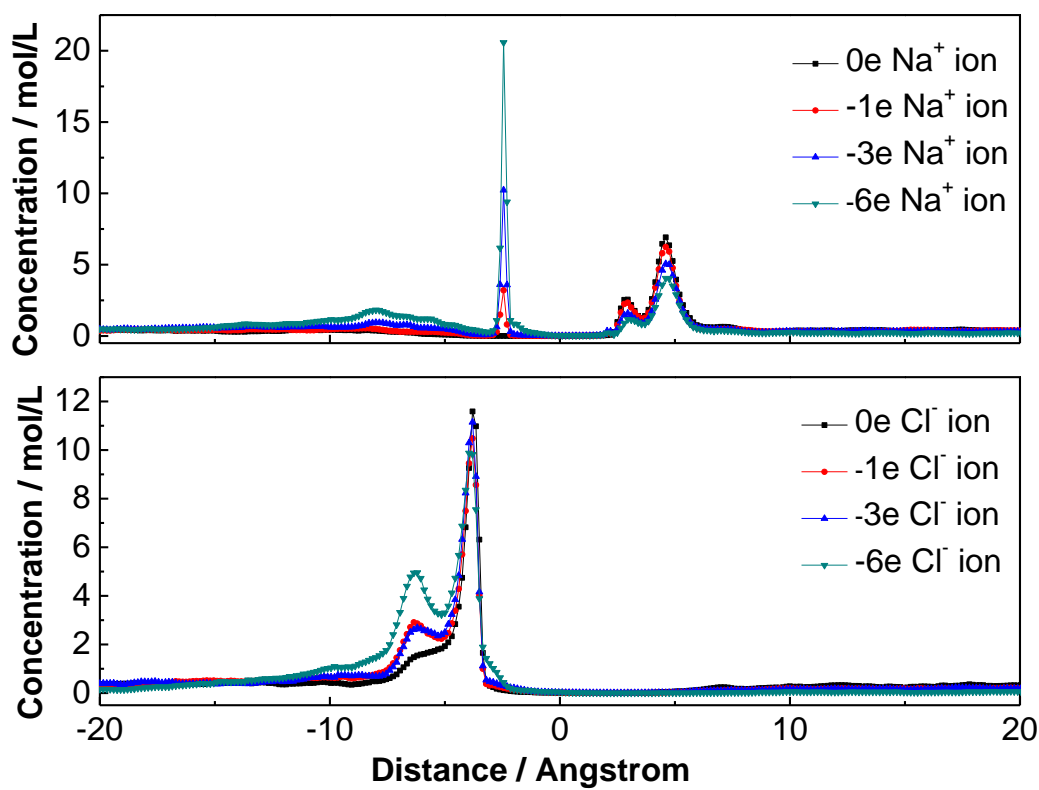

Fig.4. (color online) The ion distributions perpendicular to the graphene layers with negative charges.

When the nanopore is negatively charged, Na$^+$ ions are the selective ions. They can pass through the nanopore with priority. However, the negative pore charges have an attractive force to the ions after they pass the pore which makes another ion peak forms on the left of the graphene layer which locates at −2.4 Å. With the charge amount on the pore rim increasing, more $Na^+$ ions are adsorbed on the right side of the pores. Then, the counterion current reduces. For $Cl^-$ ions, they are non-selective to the pores. Much more $Cl^-$ ions accumulate on the left side of the graphene and a smaller peak emerges at −6.4 Å, which causes the drop of co-ion current. When the charges on the pore enhances to −6$e$, there are nearly no $Cl^-$ ions in the right part of the system.

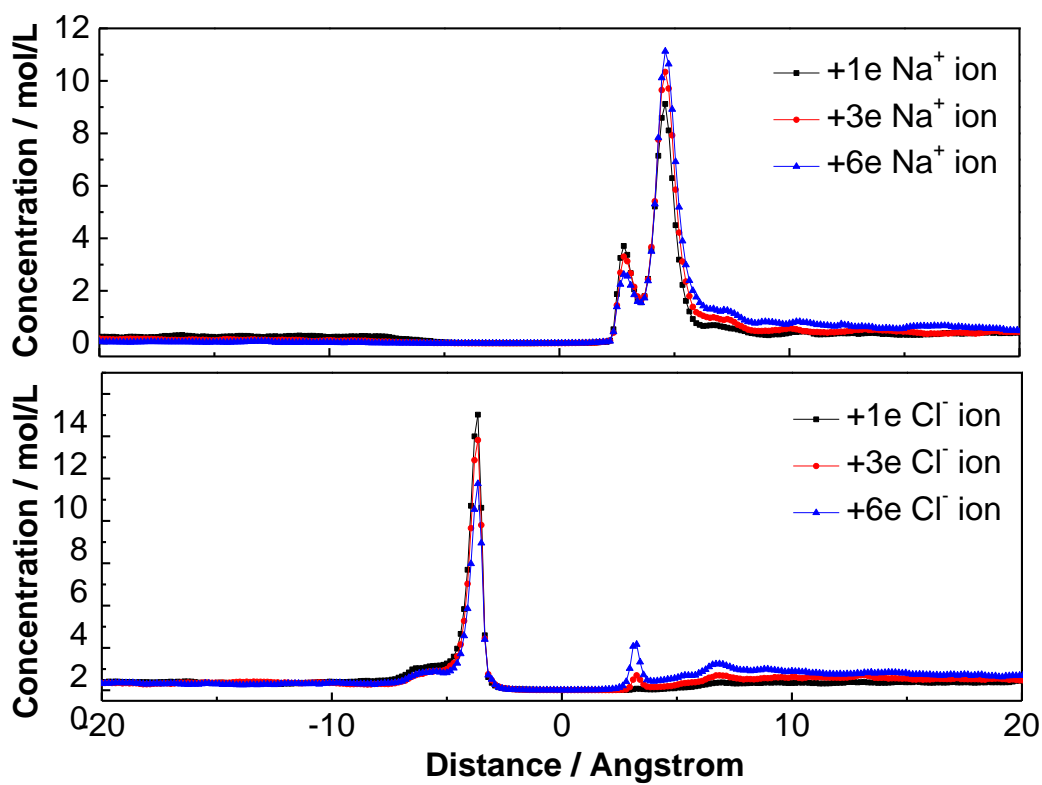

Fig.5. (color online) The ion distributions perpendicular to the graphene layer swith positive charges.

When the nanopores have positive charges, $Cl^-$ ions are the counterions and selective to the nanopores. The peak locations are 2.8 and 4.6 Å for $Na^+$ ion as well as −3.7 Å for $Cl^-$ ions as shown in Fig. 5. $Na^+$ ions accumulate on the right side of the nanopore. With the amount of charges increasing, much more $Na^+$ ions are prevented in front of the nanopore due to the rejection of the pore charges. The smaller peak reduces while the higher peak increases. While the accumulating $Cl^-$ ions on the left side of the pore decrease because of its high flux through nanopore. However, we found that fewer $Cl^-$ ions stay on the right side of the pore attracted by the charges. The ions that have passed through the nanopores distribute on the right side of the system almost uniformly which may be due to its higher mobility. This is responsible for the increased counterion current when the charges are positive, which further have effect on the water transport through the nanopores. In this system, co-ions seldom emerge on the left part when the charge amount reaches +6*e*, which is the same as the negatively charged case.

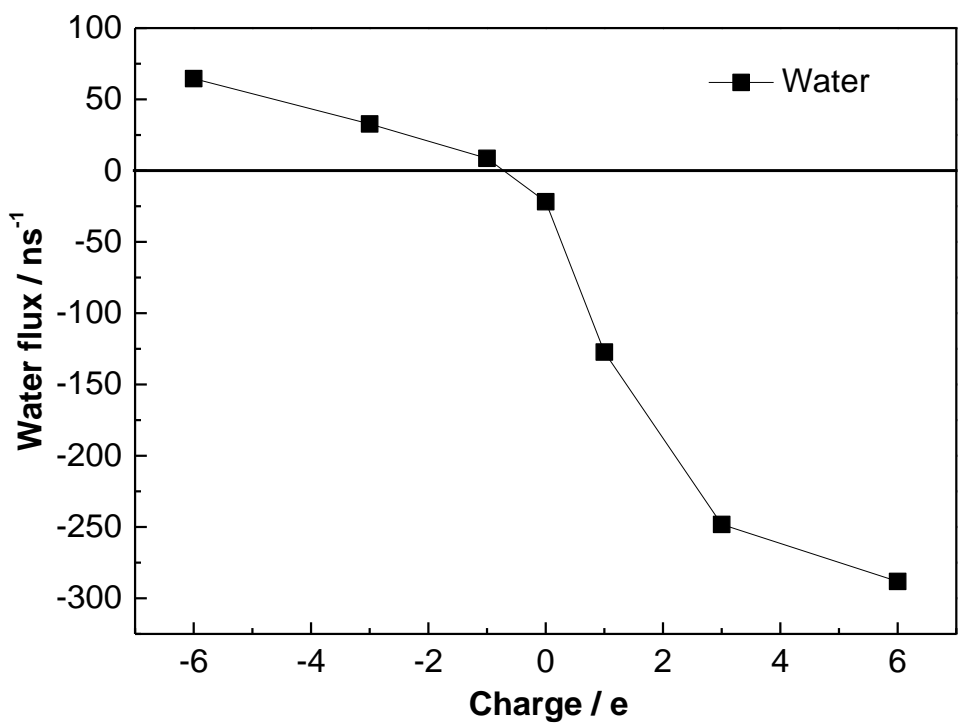

Fig. 6. The water flux in the nanopore as a function of charge amount.

Under the electric filed, water molecules have a directional movement due to the transport of ions. Fig. 6 shows the water flux in the graphene nanopore depending on the sign and amount of charges. The positive direction is set as the direction of $Na^+$ ions translocation, which is along the opposite direction of $x$ axis. It is found that the water flux can be largely enhanced by the charges on the pore rim. Besides, it increases as the charge amount rises. An interesting phenomenon is that positive charges can make water transport through nanopores much easier than negative charges. The water flux in the positive nanopore is about 10 times of that in negative pore when the charge amount is $3e$. We think this is attributed to the counterion transport through the nanopore and it is of great importance to the control of water molecule transport and sea water desalination.

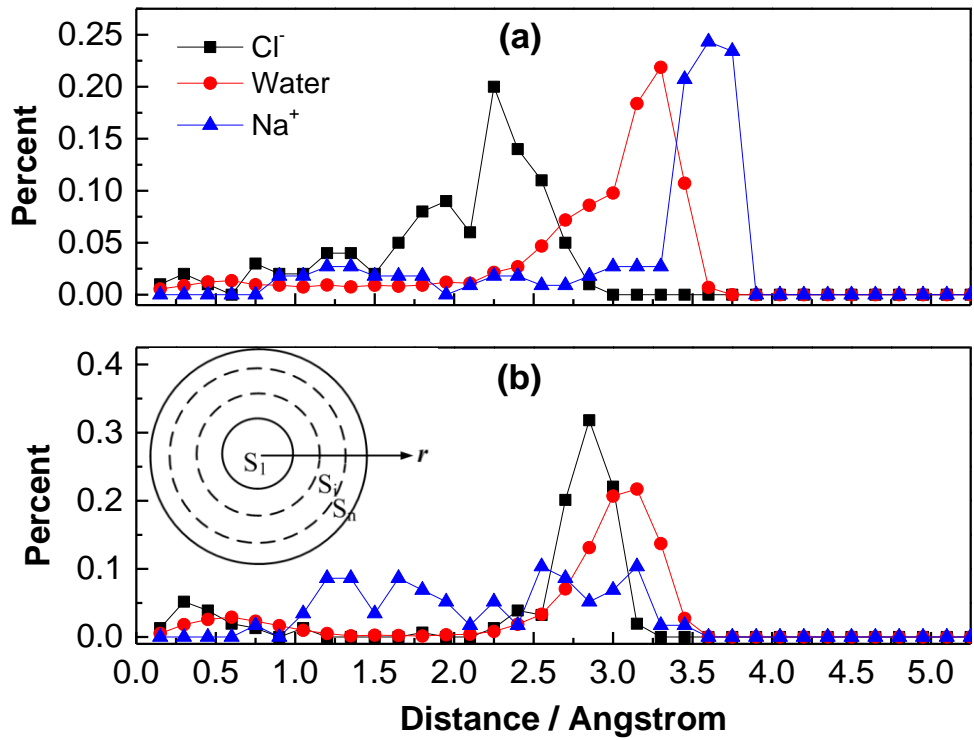

Fig. 7. (color online) The percent distributions of contributions at different positions to the total fluxes in the nanopore. (a) and (b) are obtained in negative and positive pores with charge amount $3e$, respectively.

Through analyzing the position distributions of the ions and water molecules which passed the nanopore, we obtained the percent of ions and water flux at different positions to the total fluxes in the nanopore. The binning method is shown as the inset of Fig. 7(b). The percent of each location is calculated from the following equation:

$$Percent(i) = S_i / \sum_{i=1}^{n} S_i ,$$

where $S_i$ is the number of ions or water molecules which translocate across the nanopore in the layer with number $i$. When the pore is charged with $3e$ ($e$ is the elementary charge), counterions are attracted to the surface. So the counterions in solution passing through the nanopore close to the inner surface. The main locations of $Na^+$ and $Cl^-$ ions transport are 1.65 Å and 2.45 Å from the pore surface. The farther location of $Cl^-$ ions is mainly due to their larger hydration radius.[30] In both cases, co-ions pass through the nanopore in the center of pores. While, for water molecules, the locations of the main positions are nearly the same and near the walls. As a polar molecular with isosceles triangle geometry, water molecules form hydration layers around $Na^+$ and $Cl^-$ ions. We think the water flux is mainly caused by the water molecules in the hydration layers of ions which move with the ions.

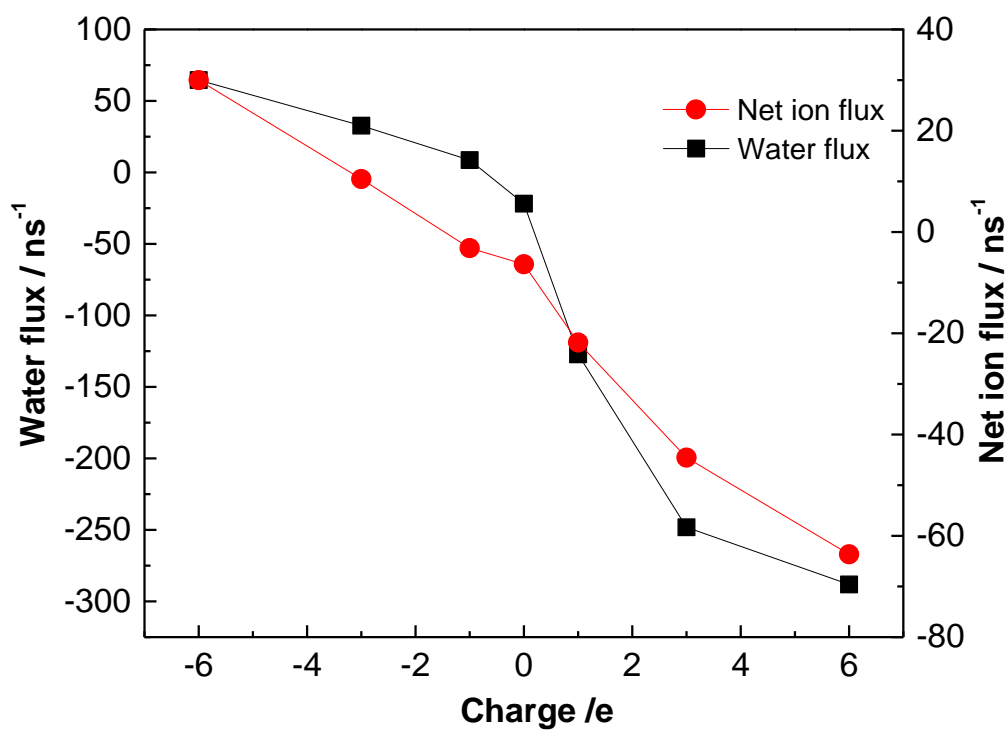

Fig. 8. (color online) The water and net ion flux depending on surface charges.

In order to confirm the direct correlation between water and ion transport across the nanopore, the net ion flux and water flux in the nanopores are shown in Fig. 8. The net ion flux is derived from the number of $Na^+$ ions which translocate through the nanopore along minus $x$ direction reducing by that of $Cl^-$ ions passing across the pore along $x$ direction. By comparing the net ion flux profile and water flux profile, we can find the two curves share similar trend. So the water flux is direct correlated to ion transport in the nanopore, especially counterion transport.

**Conclusion**

Under the same external electric filed, we investigate the ion and water transport through positively and negatively charged graphene nanopores through MD simulation. When the nanopores are charged, the counterions contribute the main current due to their passage through the nanopore with priority. The total ionic currents show an increase-decrease trend as the charge amount increase. But the counterion selection increases with the charge amount, which enhances the water flux in the nanopore, especially in positively charged pores. Because the wide application of graphene in biomolecules sensing and desalination, our simulation results can provide a clear picture of the ion and water molecule transport through charged nanopores.

Graphene is a kind of allotrope of carbon which is the main composition element of organisms. It has strong interaction with the lipid bilayers of cell membranes. Tu *et al.*[31] found that they could conduct destructive extraction of phospholipids from Escherichia coli membranes with graphene nanosheets. This is based on the strong dispersion interaction between graphene and lipid molecules. Titov *et al.*[32] simulated the sandwiched graphene-membrane superstructures and found that graphene with different layers could be hosted inside the phospholipid bilayer membrane firmly. While, Hu *et al.*[33] found that when surrounded by proteins the affinity between graphene and membrane could be largely reduced. Based on these technologies, the interaction between graphene and cell membrane can be finely controlled in the short future. We think a cell lab with a graphene gate can be sketched. On the cell, we can embed porous graphene sheet to replace a membrane region. Through modifying the edge of the pore, we can control the material exchange between the cell and outer space.